\documentclass[aps,prb,twocolumn]{revtex4-2}
\usepackage{graphicx,amsmath,amssymb,bm}

\usepackage{hyperref}
\hypersetup{
	colorlinks=true,
	linkcolor=blue,
	filecolor=magenta,      
	urlcolor=cyan,
	citecolor=red,
	anchorcolor=blue,
}

\begin{document}
	  	 		
\title{From the Gaffnian critical point to the incompressible 2/5 quantum Hall state}
\author{Sahana Das$^1$, Sudipto Das$^1$, Sutirtha Mukherjee$^2$, Sudhansu S. Mandal$^1$}
\affiliation{$^1$Department of Physics, Indian Institute of Technology, Kharagpur, West Bengal 721302, India \\
$^2$School of Physics, Korea Institute for Advanced Study, Seoul 02455, Korea	}

\date{\today}
 	
\begin{abstract}
Despite the high overlap with the exact Coulomb ground state, the so-called Gaffnian state fails to describe the incompressibility at the 2/5 quantum Hall filling factor and consequently it was conjectured to be a quantum critical state. To achieve a gapped state starting from the Gaffnian wavefunction, which we interpret as the inter-flavor pairing of the composite fermions, we propose a minimally `modified Gaffnian' wavefunction keeping the pairing intact. We find that a suitable hybridization of these two wavefunctions is an excellent description of the 2/5 quantum Hall state. It has a very high overlap with the exact Coulomb state and their entanglement spectra match up to reasonably higher levels. Interestingly, this hybridized wavefunction being a representative of a paired state suggests an exotic possibility of non-Abelian quasiparticle excitations at 2/5 filling. 
\end{abstract}
 
\maketitle

The fractional quantum Hall effect \cite{Tsui82} is one of the most fascinating cooperative phenomena of electrons because it encompasses several novel insights like Jastrow correlation \cite{Laughlin83}, flux-attached composite fermion \cite{Jain89} and boson \cite{Girvin87}, pairing \cite{Read00}, bulk-edge correspondence \cite{Wen1992,Moore91} to name a few. One of the most successful theories of fractional quantum Hall effect, namely the composite fermion (CF) theory \cite{Jain89,jain_book}, does provide an excellent description of 2/5 fractional quantum Hall state along with most of the states in the lowest Landau level. However, there was a suggestion for describing this state in terms of another wave function, which may be adiabatically connected to  Jain wave function \cite{Jain89} based on the mean-field description of the noninteracting CFs, the so-called Gaffnian wave function \cite{Simon07} derived from the conformal field theory (CFT). The Gaffnian turns out to be the zero-energy ground state of a three-body model Hamiltonian whose one part provides zero-energy ground state for the Moore-Read (MR) Pfaffian wave function \cite{Moore91} at filling factor 1/2 as a candidate for describing $\nu =5/2$ state.  
However, unlike the MR state, the corresponding CFT is non-unitary \cite{Francesco_book,Read2009b} for the Gaffnian, analogous to other paired states such as Haffnian \cite{Read00} and Haldane-Rezayi \cite{HaldaneRezayi88} states, and thus does not support gapped neutral mode of excitations \cite{Jolicoeur14,Kang17}. Some studies based on CFT even have suggested an incompatibility of this Gaffnian wave function to become the ground state wave function of a gapped state \cite{Read2009b,Freedman12}. The unsuitability of the Gaffnian wave function has further been demonstrated by the entanglement spectra (ES) \cite{Regnault09} and quasi-hole spectra \cite{Toke2009} that were inadequate to the level of consistency provided by the CF theory. Despite its high overlap with the exact Coulomb ground state, its failure has been diagnosed as lack of screening \cite{Bernevig12} and infinite correlation length \cite{Estienne15,Wu2014}.
However, there are suggestions \cite{Milovanovic2009,BoYang19,BoYang21} for transforming the non-Abelian phase of the Gaffnian to the Abelian phase of the CF theory by changing interactions or adding some background charges. 


As the Gaffnian wave function \cite{Simon07} has very high overlap with the exact Coulomb ground state yet it doesn't survive as a candidate fractional quantum Hall state, 
it was also suggested \cite{Toke2009} that 
the Gaffnian model may contain the CFs
 and that the Hilbert space of the Gaffnian model is larger in comparison to the CF model. This makes the Gaffnian state in a universality class that cannot be adiabatically connected to the CF state and possibly the Coulomb ground state. This is in sharp contrast to the conjecture \cite{Simon07} that the Gaffnian state may be a quantum critical point. In this paper, however, we show that a hybridization of the Gaffnian wave function and its modified counterpart can actually describe the 2/5 state.

 One of us has earlier formulated \cite{Mandal18_general} the lowest Landau level projected CF wave function at the filling factor 2/5 in a closed-form. In this paper, we further reformulate so that it can be directly compared with the Gaffnian wave function \cite{Simon07}. It turns out that the Gaffnian wave function is a part of the CF wave function. This immediately raises a question how the remaining subset of the CF wave function is responsible for making a gapless Gaffnian state into a gapped state? However, this subset is too complicated that hinders obtaining any suitable answer to this question. Retrospectively one may ask, is there any simpler form of wave function (formally closer to the Gaffnian function) whose linear combination with the Gaffnian wave function transforms the latter into a gapped state? The answer is affirmative, and it is proved by showing its much improved overlap with the exact state and demonstrating the corresponding entanglement spectra that have exact correspondence with the same for the Coulomb ground state up to much higher energies. In a hybrid model Hamiltonian consisting of the Gaffnian model potential and the Coulomb potential, we show that at a specific relative strength of these, the overlaps of the ground state with the Gaffnian wave function and our hybridized wave function crossover so that the latter becomes an accurate representation of the ground state for the pure Coulomb potential.

\begin{figure}[h]
	\includegraphics[width=0.7\linewidth]{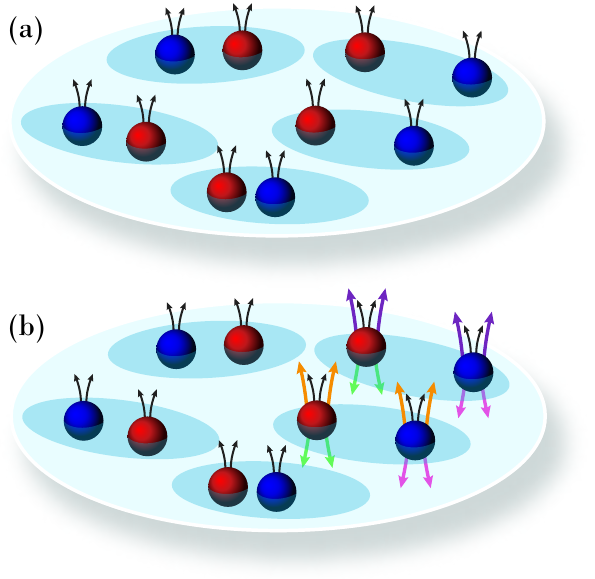}
	\caption{(Color online) (a)  A schematic representation of the Gaffnian state. While all the electrons are associated with two attached fluxes (arrows), making them composite fermions, they are divided into two flavors denoted by red and blue types. While the CFs belonging to the same flavor are noninteracting, each of them forms pair with a particle from another flavor denoted by darker cyan shaded regions. (b) A schematic representation of the modified-Gaffnian state. Only two pairs of the Gaffnian state modify their flux attachments (keeping net flux unaltered); same flavored particles mutually feel two additional negative fluxes and particles of the same pair mutually feel two additional positive fluxes. One particle can see the colored fluxes of another only if their
flux color matches.
	}
	\label{Fig:1}
\end{figure}

We begin with the interpretation of the Gaffnian wave function \cite{Simon07} 
\begin{eqnarray}
	\Psi_{\rm Gf}^{2/5}  &=& \prod_{i<j\leq N} U_{i,j}^2 \,{\cal A}\left[ \left(\prod_{k<l\leq N/2} U_{k,l}\, \,\,
	U_{k+N/2,l+N/2} \right) \right. \nonumber \\
	&& \left. \times 
	\prod_{m\leq N/2} \frac{1}{U_{m,m+N/2}}
	   \right]
	   \label{Eq:Gf}
\end{eqnarray}
in terms of the inter-flavor pairing of the CFs.
Here $N$ is the number of electrons, $U_{i,j} = z_i-z_j$, $z_j = (x_j-iy_j)/\ell$ is the complex coordinate of an electron and $\ell$ is the magnetic length, and ${\cal A}$ represents the anti-symmetrization in particle indices.
 The overall Jastrow factor $J^2=\prod_{i<j\leq N} U_{i,j}^2$ represents composite-fermionization of electrons by associating two quantum vortices to each of them. The CFs segregate themselves into two different noninteracting groups (flavors) described respectively by $J_1 = \prod_{k<l\leq N/2} U_{k,l}$ and $J_2 = \prod_{k<l\leq N/2} U_{k+N/2,l+N/2}$ which take care of the Fermi exclusion principle among particles belonging to the same flavor, followed by pairing between CFs belonging to two different flavors represented by $\prod_{m\leq N/2} U_{m,m+N/2}^{-1}$. A schematic of this paired state is shown in Fig.~\ref{Fig:1}(a). This pairing state is the exact zero-energy ground state for a certain 3-body model potential \cite{Simon07} which is given by 
\begin{equation}
	V_{3-\rm b} = \sum_{i<j<k} \left[ P_{ijk}(2Q-3) + P_{ijk}(2Q-5) \right]
	\label{Eq:3body}
\end{equation}
in spherical geometry,
where $P_{ijk}(L)$ is the projection operator that projects out the subspace of total angular momentum $L$ for three electrons, and $2Q\Phi_0$ is the total magnetic flux $(\Phi_0 = hc/e)$.

Although $\Psi_{\rm Gf}^{2/5}$ doesn't represent an incompressible state, the overlap of $\Psi_{\rm Gf}^{2/5}$ with the CF wave function  is very high in small finite systems. The reason for this is that the former is a part of the latter as follows.  
The lowest-Landau-level-projected ground state wave function for 2/5 state in the CF theory may be written in a closed-form (see Ref.\onlinecite{Mandal18_general} for details) as
\begin{eqnarray}
	\Psi^{2/5}_{\rm CF} &=& \prod_{i<j\leq N} U_{i,j}^2 \,{\cal A} \left[\left(\prod_{k<l\leq N/2} U_{k,l}\, 
	U_{k+N/2,l+N/2} \right) \right. \nonumber \\
	&&\times \left. \prod_{m\leq N/2} \left( \sum_{k=1,\neq m}^N \frac{1}{U_{m,k}} \right)	\right]
	\label{Eq:CF}
\end{eqnarray}
It clearly shows that $ \Psi^{2/5}_{\rm Gf}$ is a part of $\Psi^{2/5}_{\rm CF}$. As the correlation-physics of these two wave functions are entirely different, the former as a pairing state between two different flavors of CFs whereas the latter corresponds to the mean-field description of non-interacting CFs filling two effective Landau levels, it can not be readily understood how the remaining part of the $\Psi_{\rm CF}^{2/5}$ wave function can make a compressible state incompressible or how a non-unitary CFT transforms into a unitary CFT.

The above issue motivates us to search for a suitable simple function whose amalgamation with the Gaffnian can make the state incompressible. 
To this end, we propose an additional wave function 
\begin{eqnarray}
	&&\Psi_{\rm m-Gf}^{2/5}  = \prod_{i<j\leq N} U_{i,j}^2 \,{\cal A}\left[\left(\prod_{k<l\leq N/2} U_{k,l}\, \,\,
	U_{k+N/2,l+N/2} \right) \right. \nonumber \\
	& \times&  \left. \left(	\prod_{m\leq N/2} \frac{1}{U_{m,m+N/2}} \right) \sum_{l_1< l_2\leq N/2}\left(
	\frac{U_{l_1,l_1+N/2}^2 U_{l_2,l_2+N/2}^2}{U_{l_1,l_2}^{2} U_{l_1+N/2,l_2+N/2}^{2}} 
	\right)  \right] \nonumber \\ \label{Eq:mGf}	   
\end{eqnarray}
which is minimally modified from the Gaffnian wave function and hence named as modified-Gaffnian wave function.
This modified-Gaffnian wave function may be interpreted as follows. While $(N/2-2)$ pairs of CFs of dissimilar flavors in the Gaffnian state remain unaltered, the change occurs (see a schematic in Fig.~\ref{Fig:1}b) for the remaining two pairs in terms of the attachment of additional flux which are invisible to other $(N/2-2)$ pairs. The CFs with the same flavor in these two different pairs (different flavors of the same pair) feel two less (more) zeros at each other's place; in other words, find two additional negative (positive) flux attachments. However, neither total flux felt by each CF changes, nor the pair breaking occurs. Consequently, like $\psi_{\rm CF}^{2/5}$ and $\Psi_{\rm Gf}^{2/5}$, the highest accessible single-particle orbital in $\Psi_{\rm m-Gf}^{2/5}$ is also $5N/2 -4$.
We note that $\Psi_{\rm m-Gf}^{2/5}$ doesn't belong to the remaining   subset (excess of the Gaffnian subset) of $\Psi_{\rm CF}^{2/5}$ discussed above.

\begin{figure}[h]
	\includegraphics[width=\linewidth]{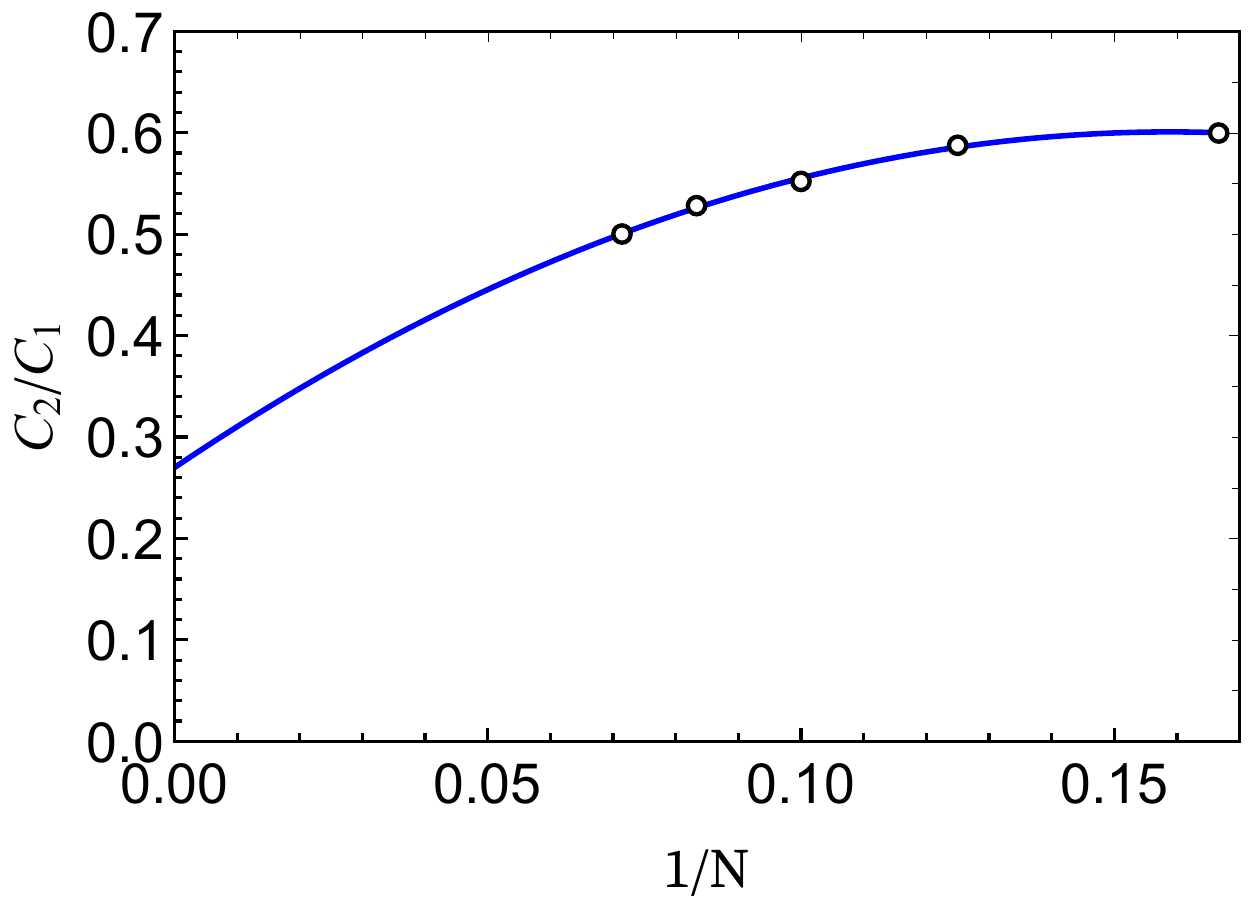}
	\caption{(Color online) Representation of the ratio $C_2/C_1$ from Eq.~\ref{Eq:Hybrid} {\it versus} $1/N$. 
	The extrapolation of quadratic fit suggests that the ratio is finite even in the thermodynamic limit.  }
	\label{Fig:2}
\end{figure}

We next consider the hybridization of  $\Psi_{\rm m-Gf}^{2/5}$ and $\Psi_{\rm Gf}^{2/5}$ as
\begin{equation}
	\Psi_{\rm Hybrid}^{2/5} = C_1 \tilde{\Psi}_{\rm Gf}^{2/5}+C_2 \tilde{\Psi}_{\rm m-Gf}^{2/5}
	\label{Eq:Hybrid}
\end{equation}
where $\tilde{\Psi}_{\rm Gf}^{2/5}$ and $\tilde{\Psi}_{\rm m-Gf}^{2/5}$ are the normalized form of $\Psi_{\rm Gf}^{2/5}$ and $\Psi_{\rm m-Gf}^{2/5}$ respectively.
The ratio of the coefficients $C_1$ and $C_2$ 
is optimized for maximum overlap of $\Psi_{\rm Hybrid}^{2/5}$ with the exact ground state for the Coulomb potential $V_c = e^2/\epsilon r$ in the spherical geometry. 
By the stereographic projection \cite{HaldaneSphere}, the wave functions $\Psi_{\rm Gf}^{2/5}$, $\Psi_{\rm m-Gf}^{2/5}$ and $\Psi_{\rm Hybrid}^{2/5}$ are transformed for the spherical geometry with $U_{ij} = u_iv_j-v_iu_j$ where $u_j = \cos(\theta_j/2) e^{i\phi_j/2}$ and $v_j = \sin (\theta_j/2)e^{-i\phi_j/2}$ are the spherical spinors with spherical angles $0\leq \theta_j \leq \pi$ and $0\leq \phi_j \leq 2\pi$. 
A thermodynamic extrapolation of $C_2/C_1$ shown in Fig.~\ref{Fig:2} suggests that the hybridization cannot be ignored even in the thermodynamic limit. 
The overlap of various wave functions, including the exact Coulomb ground state $\Psi_{\rm Ex}^{2/5}$, are tabulated in Table~\ref{Table:1}. The wave function $\Psi_{\rm Hybrid}^{2/5}$ is found to be an extremely accurate description of the exact state and its overlap with $\Psi_{\rm CF}^{2/5}$ is also very high. The Coulomb energies corresponding to the states $\Psi_{\rm CF}^{2/5}$, $\Psi_{\rm Gf}^{2/5}$, $\Psi_{\rm Hybrid}^{2/5}$ and the exact state are compared in Table~\ref{Table:2}. The ground state energy for $\Psi_{\rm Hybrid}^{2/5}$ is as accurate as the same for $\Psi_{\rm CF}^{2/5}$.

\begin{widetext}

\begin{table}[h]
	\centering
	\caption{Overlap of the different states at $\nu = 2/5$ up to 14 electrons in spherical geometry. The numbers inside parentheses indicate the corresponding statistical errors in the last significant digit due to evaluation by the Metropolis Monte Carlo method. Here $\Psi_{\rm CF}^{2/5}$ has been taken as the original form of CF wave function in spherical geometry \cite{jain_book} (not the stereographic projected form of $\Psi_{\rm CF}^{2/5}$ in Eq.~\ref{Eq:CF}). The values without parenthesis are exact as they are evaluated by the method of decomposition into single particle eigen basis (DSPEB) \cite{Das21}. The DSPEB couldn't be performed for the composite fermion wave function and for other wave functions with a larger number of particles.}
	\begin{tabular}{c|c|c|c|c|c|c|c} \hline\hline
		N & $\langle \Psi^{2/5}_{\rm Ex}|\Psi^{2/5}_{\rm CF} \rangle$ & $\langle \Psi^{2/5}_{\rm Ex}|\Psi^{2/5}_{\rm Gf} \rangle$ & $\langle \Psi^{2/5}_{\rm Ex}|\Psi^{2/5}_{\rm m-Gf} \rangle$ & $\langle \Psi^{2/5}_{\rm Ex}|\Psi^{2/5}_{\rm Hybrid} \rangle$ & $\langle \Psi^{2/5}_{\rm CF}|\Psi^{2/5}_{\rm Gf} \rangle$ & $\langle \Psi^{2/5}_{\rm CF}|\Psi^{2/5}_{\rm m-Gf} \rangle$ & $\langle \Psi^{2/5}_{\rm CF}|\Psi^{2/5}_{\rm Hybrid} \rangle$  \\ \hline

        6 & 0.999687 (1) & 0.9883947 & 0.9676911 & 0.999831  & 0.99100 (1) & 0.9720 (0) & 0.999855 (0) \\	\hline
        
        8 & 0.999372 (3) & 0.9771221 & 0.9327855 & 0.999709 & 0.98250 (5) & 0.9419 (2) & 0.999665 (1)\\	\hline
        
        10 & 0.997779 (4) & 0.9715446 & 0.9098972 & 0.997357 & 0.97724 (6) & 0.9182 (2) & 0.999578 (2) \\	\hline
        
        12 & 0.99689 (1) & 0.9644 (2) & 0.8789 (5) & 0.99555 (2) & 0.9724 (1) & 0.8907 (4) & 0.998936 (6) \\	\hline
        
        14 & 0.99573 (2) & 0.9584 (3) & 0.8463 (7) & 0.99296 (5) & 0.9683 (2) & 0.8605 (6) & 0.99813 (2) \\	\hline


	\end{tabular}
	\label{Table:1}
\end{table}

\end{widetext}

\begin{table}[h]
	\caption{Comparison between exact Coulomb ground state energy and the energies corresponding to the wave functions $\Psi_{\rm CF}^{2/5}$ (original form), $\Psi_{\rm Gf}^{2/5}$, and $\Psi_{\rm Hybrid}^{2/5}$. The energies are in the unit of $e^2/\epsilon\ell$ and per electron (after subtracting the background energies).} 
	\begin{tabular}{c|c|c|c|c}\hline\hline
		N & $ E_{\rm Ex}$ & $E_{\rm CF}$ & $E_{\rm Gf}$ &  $E_{\rm Hybrid} $  \\ \hline 
		
		6 & -0.50040  & -0.5004 (2) &	-0.4994 (2)  & -0.5004 (2) \\
		
		8  & -0.48024 & -0.4802 (3) & -0.4788 (4) &  -0.4802 (3)  \\
		
		10 & -0.46945 & -0.4693 (5) & -0.4680 (6)  & -0.4693 (5)   \\
		
		12 & -0.46265 & -0.4625 (7) & -0.4611 (9)  & -0.4625 (7)  \\
		
		14 & -0.45799 & -0.458 (2) & -0.457 (2)  & -0.458 (2) \\
		
		\hline
	\end{tabular}
\label{Table:2}
\end{table}

\begin{figure}[h]	
	\includegraphics[width=\linewidth]{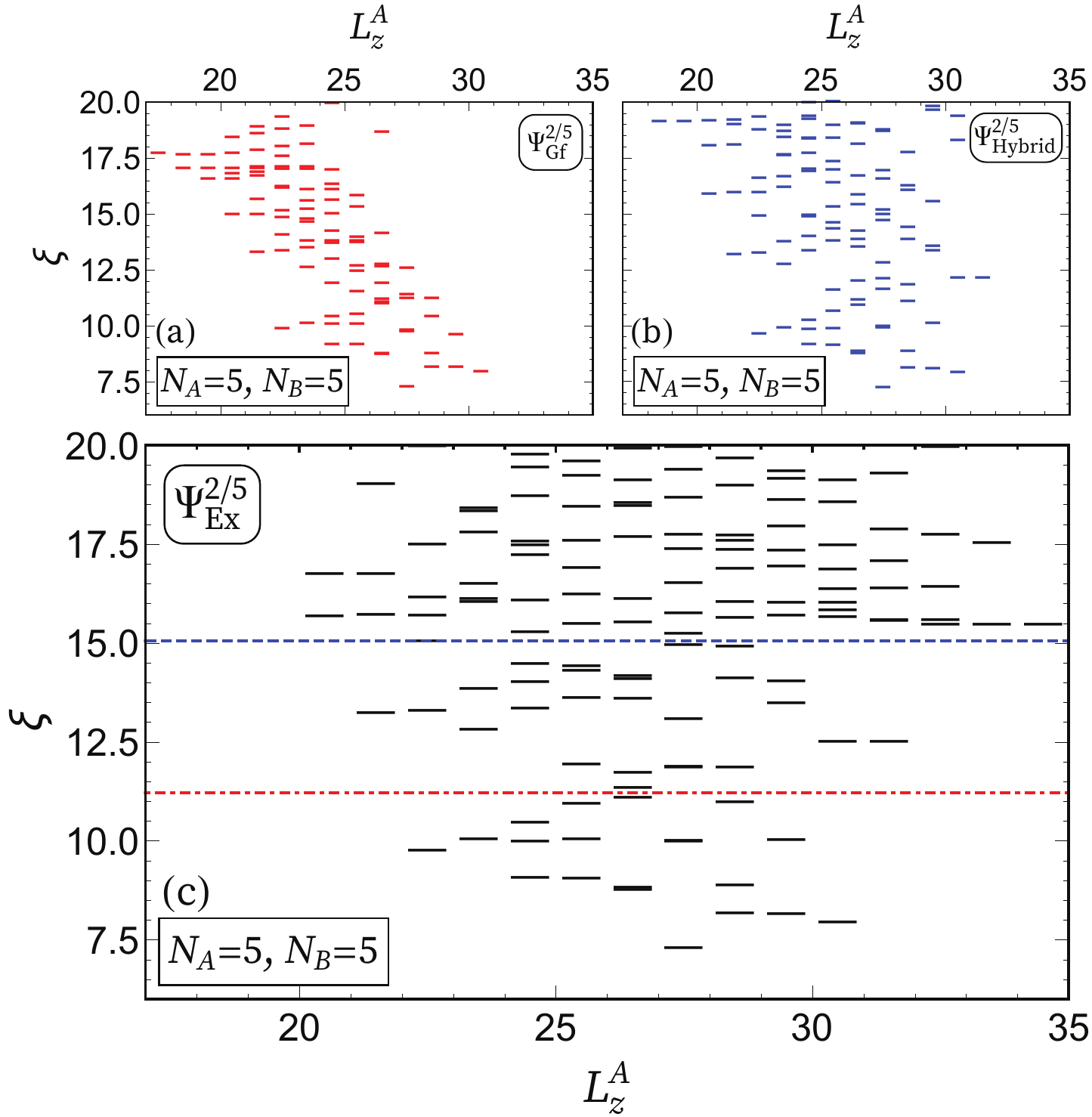}
	\caption{(Color online) Topological entanglement spectra at filling factor $\nu=2/5$ for 10 particle system taking $N_A=N_B=5$ for (a) Gaffnian wave function, (b) Hybridized wave function, and (c) Exact wave function. In the bottom panel, the levels below the red dash-dotted line are almost identical to the Gaffnian levels and the levels below the blue dashed line are identical to those of our Hybridized wave function. Here $L_z^A$ is the sum of the angular momenta of the occupied orbitals in the upper hemisphere. }
	\label{Fig:3}
\end{figure}

We next determine the entanglement spectra (ES)\cite{Regnault09,LiHaldane2008,Simon12,Sterdyniak12} by the method of particle partition between upper and lower hemispheres in the spherical geometry.
Recall that the ES for $\Psi_{\rm Gf}$ was insufficient \cite{Regnault09} baring very low-lying sector for reproducing the ES for the exact state.
 Figure \ref{Fig:3}(a) and \ref{Fig:3}(b) show the ES for $\Psi_{\rm Gf}^{2/5}$ and $\Psi_{\rm Hybrid}^{2/5}$ respectively and are compared with the ES for the exact state in Fig.~\ref{Fig:3}(c). The wave function  $\Psi_{\rm Hybrid}^{2/5}$ reproduces the ES up to sufficiently higher energy. 
 Therefore, $\Psi_{\rm Hybrid}^{2/5}$ not only has very high overlap with the Coulomb ground state and reproduces almost accurate ground state energy, but it also exhibits similar topological order as the exact state as far as the ES is concerned.

 Though the energies of both $\Psi_{\rm CF}^{2/5}$ and $\Psi_{\rm Hybrid}^{2/5}$ are very close to the exact groundstate energy, the construction of these wavefunctions are however entirely different.
 The former is believed to support Abelian quasiparticles, whereas the latter being a paired state of the CFs is very likely to support \cite{Read00} non-Abelian quasiparticles. This suggests that there may be two competing topological orders in 2/5 quantum Hall state.

\begin{figure}[h]
	\includegraphics[width=\linewidth]{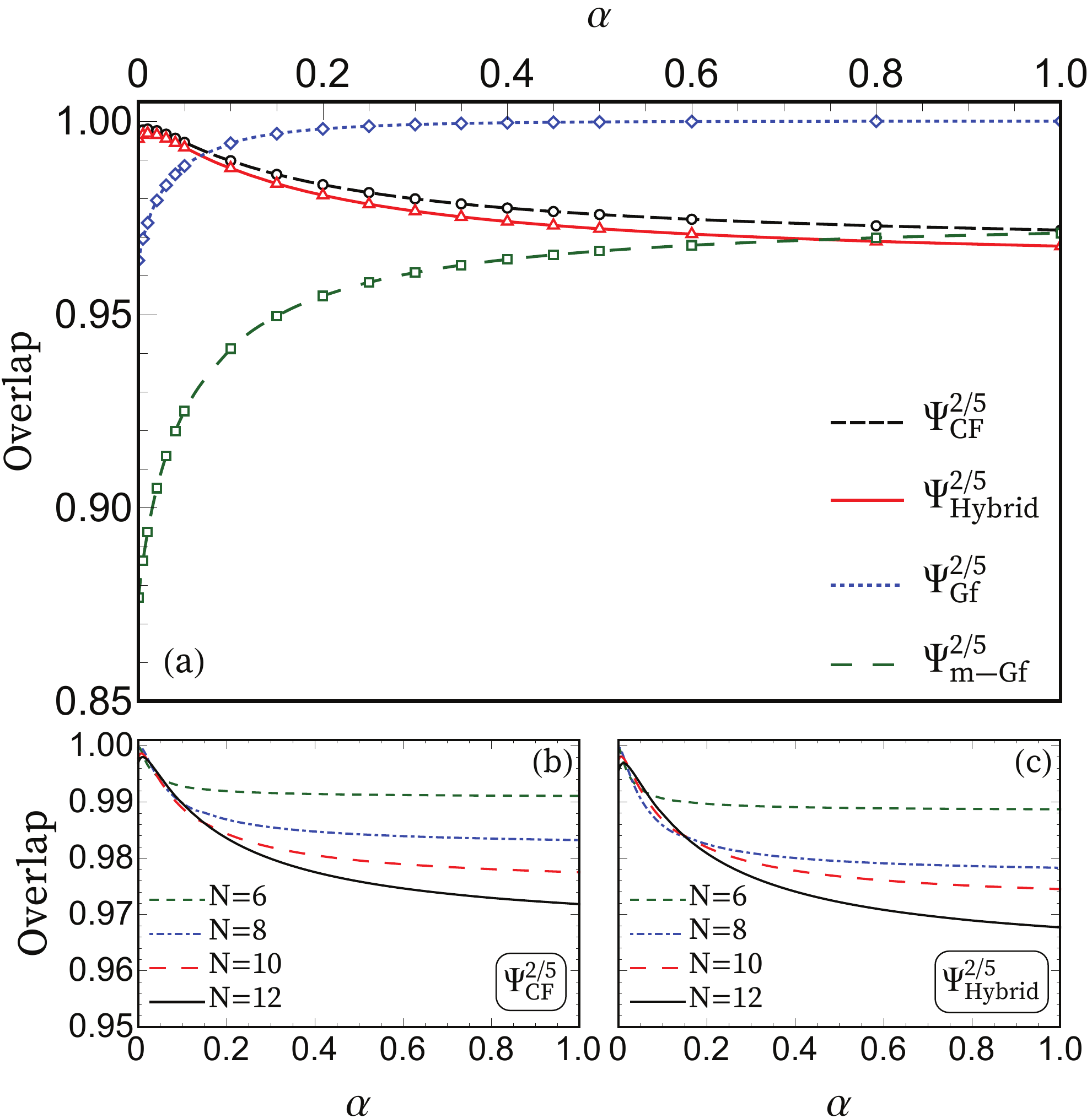}
	\caption{(Color online) (a) Overlap of $\Psi_{\rm CF}^{2/5}$, $\Psi_{\rm Hybrid}^{2/5}$, $\Psi_{\rm Gf}^{2/5}$, and $\Psi_{\rm m-Gf}^{2/5}$ with the ground state of model potential $H_\alpha$,  $\Psi_\alpha^{2/5}$ for N=12. (b,c) Overlap of $\Psi_{\rm CF}^{2/5}$ and $\Psi_{\rm Hybrid}^{2/5}$ with $\Psi_\alpha^{2/5}$ respectively for $N=6,\, 8,\, 10$, and $12$.}
	\label{Fig:4}
\end{figure}

Having shown that $\Psi_{\rm Hybrid}^{2/5}$, a hybridized state of the Gaffnian and the modified-Gaffnian wave function, is an excellent ground state wave function, we perform a case study of how it is favored over the Gaffnian as the strength of the two-body Coulomb interaction is increased starting from the full $V_{3-\rm b}$.	Performing exact diagonalization of a mixed Hamiltonian \cite{Toke2009,Jolicoeur14}
\begin{equation}
	H_{\rm \alpha} = \alpha V_{3-\rm b} + (1-\alpha)V_C 
	\label{eq:Hmix}
\end{equation}
and varying the parameter $0\leq \alpha \leq 1$, we calculate and show overlaps of the corresponding ground state with $\Psi_{\rm CF}^{2/5}$,  $\Psi_{\rm Hybrid}^{2/5}$, $\Psi_{\rm Gf}^{2/5}$, and $\Psi_{\rm m-Gf}^{2/5}$ wave functions in Fig.~\ref{Fig:4}(a). As expected, for higher $\alpha$, $\Psi_{\rm Gf}^{2/5}$ is an excellent ground state while all other states are unsuitable for capturing the 3-body nature of the interacting potential. However, the overlaps of both $\Psi_{\rm Gf}^{2/5}$ and $\Psi_{\rm m-Gf}^{2/5}$ significantly decrease as the two-body interaction limit is approached. On the other hand, the overlap of $\Psi_{\rm Hybrid}^{2/5}$ with the exact ground state approaches to unity. The crossover of the overlap for the Gaffnian and the hybridized state occurs at $\alpha \approx 0.07$ for $N=12$.
We note that the overlaps for the hybridized wave function and the CF wave function are almost identical for all values of $\alpha$ and for different $N$ (Figs.~\ref{Fig:4}b and c). 
Since the hybridized state  has the highest overlap for $\alpha$ close to the Coulomb point, we may consider the corresponding model Hamiltonian ${H_\alpha}$ ($\alpha \lesssim 0.07$) as roughly the parent Hamiltonian for the hybridized state. As the extrapolated gap for such a Hamiltonian does not close \cite{Jolicoeur14} in the thermodynamic limit, the hybridized wave function is a description of an incompressible state.

There are discrepancies found \cite{Toke2009} in the characteristics of the excitation spectra between the Gaffnian state and the composite fermion wave function  as well as the exact state.  
Also, the diverging correlation length \cite{Estienne15,Wu2014} of the Gaffnian state gives a signal of gaplessness. It will therefore be necessarily important to check these properties for the hybridized wave function proposed here along with its screening behavior \cite{Bernevig12}. Some of these will be taken up in a future publication. A demonstration of the transformation from the non-unitary CFT of the Gaffnian state to the presumably unitary CFT of the hybridized state should also be an important direction of future study. 
As the hybridized wave function is an outcome of the pairing between composite fermions of dissimilar flavors, a fractional thermal Hall conductance \cite{Banerjee2018} is expected in 2/5 state. Also, as per general belief, a paired state will support non-Abelian quasiparticles and hence it is extremely important to study the braiding statistics \cite{Halperin21} of the quasiparticles at the filling factor 2/5. Under what conditions the Abelian phase of the composite fermion theory or the non-Abelian phase of the hybridized paired state will occur should also be an interesting question to investigate. As the energies of both the phases are extremely close, there could be a spontaneous possession of one of these topological orders based on the external conditions.


We thank Thierry Jolicoeur and Csaba T\ifmmode \mbox{\H{o}}\else \H{o}\fi{}ke  for providing their respective raw data of their publications related to the mixing of three-body and two-body potentials. It is a pleasure to acknowledge the valuable comments of Csaba T\ifmmode \mbox{\H{o}}\else \H{o}\fi{}ke and Bo Yang on the initial post of the manuscript in the arXiv. We thank the developers of the DiagHam package for keeping it open access. We acknowledge the Param Shakti (IIT Kharagpur)–a National Supercomputing Mission, Government of India for providing their computational resources. SSM is supported by SERB through its grant MTR/2019/000546. SM is supported by KIAS individual grant PG034303.

\bibliographystyle{apsrev4-2}
\bibliography{reference_25.bib}

\begin{thebibliography}{31}%
\makeatletter
\providecommand \@ifxundefined [1]{%
 \@ifx{#1\undefined}
}%
\providecommand \@ifnum [1]{%
 \ifnum #1\expandafter \@firstoftwo
 \else \expandafter \@secondoftwo
 \fi
}%
\providecommand \@ifx [1]{%
 \ifx #1\expandafter \@firstoftwo
 \else \expandafter \@secondoftwo
 \fi
}%
\providecommand \natexlab [1]{#1}%
\providecommand \enquote  [1]{``#1''}%
\providecommand \bibnamefont  [1]{#1}%
\providecommand \bibfnamefont [1]{#1}%
\providecommand \citenamefont [1]{#1}%
\providecommand \href@noop [0]{\@secondoftwo}%
\providecommand \href [0]{\begingroup \@sanitize@url \@href}%
\providecommand \@href[1]{\@@startlink{#1}\@@href}%
\providecommand \@@href[1]{\endgroup#1\@@endlink}%
\providecommand \@sanitize@url [0]{\catcode `\\12\catcode `\$12\catcode
  `\&12\catcode `\#12\catcode `\^12\catcode `\_12\catcode `\%12\relax}%
\providecommand \@@startlink[1]{}%
\providecommand \@@endlink[0]{}%
\providecommand \url  [0]{\begingroup\@sanitize@url \@url }%
\providecommand \@url [1]{\endgroup\@href {#1}{\urlprefix }}%
\providecommand \urlprefix  [0]{URL }%
\providecommand \Eprint [0]{\href }%
\providecommand \doibase [0]{https://doi.org/}%
\providecommand \selectlanguage [0]{\@gobble}%
\providecommand \bibinfo  [0]{\@secondoftwo}%
\providecommand \bibfield  [0]{\@secondoftwo}%
\providecommand \translation [1]{[#1]}%
\providecommand \BibitemOpen [0]{}%
\providecommand \bibitemStop [0]{}%
\providecommand \bibitemNoStop [0]{.\EOS\space}%
\providecommand \EOS [0]{\spacefactor3000\relax}%
\providecommand \BibitemShut  [1]{\csname bibitem#1\endcsname}%
\let\auto@bib@innerbib\@empty
\bibitem [{\citenamefont {Tsui}\ \emph {et~al.}(1982)\citenamefont {Tsui},
  \citenamefont {Stormer},\ and\ \citenamefont {Gossard}}]{Tsui82}%
  \BibitemOpen
  \bibfield  {author} {\bibinfo {author} {\bibfnamefont {D.~C.}\ \bibnamefont
  {Tsui}}, \bibinfo {author} {\bibfnamefont {H.~L.}\ \bibnamefont {Stormer}},\
  and\ \bibinfo {author} {\bibfnamefont {A.~C.}\ \bibnamefont {Gossard}},\
  }\href {https://doi.org/10.1103/PhysRevLett.48.1559} {\bibfield  {journal}
  {\bibinfo  {journal} {Phys. Rev. Lett.}\ }\textbf {\bibinfo {volume} {48}},\
  \bibinfo {pages} {1559} (\bibinfo {year} {1982})}\BibitemShut {NoStop}%
\bibitem [{\citenamefont {Laughlin}(1983)}]{Laughlin83}%
  \BibitemOpen
  \bibfield  {author} {\bibinfo {author} {\bibfnamefont {R.~B.}\ \bibnamefont
  {Laughlin}},\ }\href {https://doi.org/10.1103/PhysRevLett.50.1395} {\bibfield
   {journal} {\bibinfo  {journal} {Phys. Rev. Lett.}\ }\textbf {\bibinfo
  {volume} {50}},\ \bibinfo {pages} {1395} (\bibinfo {year}
  {1983})}\BibitemShut {NoStop}%
\bibitem [{\citenamefont {Jain}(1989)}]{Jain89}%
  \BibitemOpen
  \bibfield  {author} {\bibinfo {author} {\bibfnamefont {J.~K.}\ \bibnamefont
  {Jain}},\ }\href {https://doi.org/10.1103/PhysRevLett.63.199} {\bibfield
  {journal} {\bibinfo  {journal} {Phys. Rev. Lett.}\ }\textbf {\bibinfo
  {volume} {63}},\ \bibinfo {pages} {199} (\bibinfo {year} {1989})}\BibitemShut
  {NoStop}%
\bibitem [{\citenamefont {Girvin}\ and\ \citenamefont
  {MacDonald}(1987)}]{Girvin87}%
  \BibitemOpen
  \bibfield  {author} {\bibinfo {author} {\bibfnamefont {S.~M.}\ \bibnamefont
  {Girvin}}\ and\ \bibinfo {author} {\bibfnamefont {A.~H.}\ \bibnamefont
  {MacDonald}},\ }\href {https://doi.org/10.1103/PhysRevLett.58.1252}
  {\bibfield  {journal} {\bibinfo  {journal} {Phys. Rev. Lett.}\ }\textbf
  {\bibinfo {volume} {58}},\ \bibinfo {pages} {1252} (\bibinfo {year}
  {1987})}\BibitemShut {NoStop}%
\bibitem [{\citenamefont {Read}\ and\ \citenamefont {Green}(2000)}]{Read00}%
  \BibitemOpen
  \bibfield  {author} {\bibinfo {author} {\bibfnamefont {N.}~\bibnamefont
  {Read}}\ and\ \bibinfo {author} {\bibfnamefont {D.}~\bibnamefont {Green}},\
  }\href {https://doi.org/10.1103/PhysRevB.61.10267} {\bibfield  {journal}
  {\bibinfo  {journal} {Phys. Rev. B}\ }\textbf {\bibinfo {volume} {61}},\
  \bibinfo {pages} {10267} (\bibinfo {year} {2000})}\BibitemShut {NoStop}%
\bibitem [{\citenamefont {Wen}(1992)}]{Wen1992}%
  \BibitemOpen
  \bibfield  {author} {\bibinfo {author} {\bibfnamefont {X.-G.}\ \bibnamefont
  {Wen}},\ }\href@noop {} {\bibfield  {journal} {\bibinfo  {journal}
  {International Journal of Modern Physics B}\ }\textbf {\bibinfo {volume}
  {06}},\ \bibinfo {pages} {1711} (\bibinfo {year} {1992})}\BibitemShut
  {NoStop}%
\bibitem [{\citenamefont {Moore}\ and\ \citenamefont {Read}(1991)}]{Moore91}%
  \BibitemOpen
  \bibfield  {author} {\bibinfo {author} {\bibfnamefont {G.}~\bibnamefont
  {Moore}}\ and\ \bibinfo {author} {\bibfnamefont {N.}~\bibnamefont {Read}},\
  }\href {https://doi.org/https://doi.org/10.1016/0550-3213(91)90407-O}
  {\bibfield  {journal} {\bibinfo  {journal} {Nuclear Physics B}\ }\textbf
  {\bibinfo {volume} {360}},\ \bibinfo {pages} {362 } (\bibinfo {year}
  {1991})}\BibitemShut {NoStop}%
\bibitem [{\citenamefont {Jain}(2007)}]{jain_book}%
  \BibitemOpen
  \bibfield  {author} {\bibinfo {author} {\bibfnamefont {J.~K.}\ \bibnamefont
  {Jain}},\ }\href {https://doi.org/10.1017/CBO9780511607561} {\emph {\bibinfo
  {title} {Composite Fermions}}}\ (\bibinfo  {publisher} {Cambridge University
  Press},\ \bibinfo {year} {2007})\BibitemShut {NoStop}%
\bibitem [{\citenamefont {Simon}\ \emph {et~al.}(2007)\citenamefont {Simon},
  \citenamefont {Rezayi}, \citenamefont {Cooper},\ and\ \citenamefont
  {Berdnikov}}]{Simon07}%
  \BibitemOpen
  \bibfield  {author} {\bibinfo {author} {\bibfnamefont {S.~H.}\ \bibnamefont
  {Simon}}, \bibinfo {author} {\bibfnamefont {E.~H.}\ \bibnamefont {Rezayi}},
  \bibinfo {author} {\bibfnamefont {N.~R.}\ \bibnamefont {Cooper}},\ and\
  \bibinfo {author} {\bibfnamefont {I.}~\bibnamefont {Berdnikov}},\ }\href
  {https://doi.org/10.1103/PhysRevB.75.075317} {\bibfield  {journal} {\bibinfo
  {journal} {Phys. Rev. B}\ }\textbf {\bibinfo {volume} {75}},\ \bibinfo
  {pages} {075317} (\bibinfo {year} {2007})}\BibitemShut {NoStop}%
\bibitem [{\citenamefont {Francesco}\ \emph {et~al.}(1997)\citenamefont
  {Francesco}, \citenamefont {Mathieu},\ and\ \citenamefont
  {S\'en\'echal}}]{Francesco_book}%
  \BibitemOpen
  \bibfield  {author} {\bibinfo {author} {\bibfnamefont {P.~D.}\ \bibnamefont
  {Francesco}}, \bibinfo {author} {\bibfnamefont {P.}~\bibnamefont {Mathieu}},\
  and\ \bibinfo {author} {\bibfnamefont {D.}~\bibnamefont {S\'en\'echal}},\
  }\href@noop {} {\emph {\bibinfo {title} {Conformal Field Theory}}}\ (\bibinfo
   {publisher} {Springer-Verlag},\ \bibinfo {year} {1997})\BibitemShut
  {NoStop}%
\bibitem [{\citenamefont {Read}(2009)}]{Read2009b}%
  \BibitemOpen
  \bibfield  {author} {\bibinfo {author} {\bibfnamefont {N.}~\bibnamefont
  {Read}},\ }\href {https://doi.org/10.1103/PhysRevB.79.245304} {\bibfield
  {journal} {\bibinfo  {journal} {Phys. Rev. B}\ }\textbf {\bibinfo {volume}
  {79}},\ \bibinfo {pages} {245304} (\bibinfo {year} {2009})}\BibitemShut
  {NoStop}%
\bibitem [{\citenamefont {Haldane}\ and\ \citenamefont
  {Rezayi}(1988)}]{HaldaneRezayi88}%
  \BibitemOpen
  \bibfield  {author} {\bibinfo {author} {\bibfnamefont {F.~D.~M.}\
  \bibnamefont {Haldane}}\ and\ \bibinfo {author} {\bibfnamefont {E.~H.}\
  \bibnamefont {Rezayi}},\ }\href {https://doi.org/10.1103/PhysRevLett.60.956}
  {\bibfield  {journal} {\bibinfo  {journal} {Phys. Rev. Lett.}\ }\textbf
  {\bibinfo {volume} {60}},\ \bibinfo {pages} {956} (\bibinfo {year}
  {1988})}\BibitemShut {NoStop}%
\bibitem [{\citenamefont {Jolicoeur}\ \emph {et~al.}(2014)\citenamefont
  {Jolicoeur}, \citenamefont {Mizusaki},\ and\ \citenamefont
  {Lecheminant}}]{Jolicoeur14}%
  \BibitemOpen
  \bibfield  {author} {\bibinfo {author} {\bibfnamefont {T.}~\bibnamefont
  {Jolicoeur}}, \bibinfo {author} {\bibfnamefont {T.}~\bibnamefont
  {Mizusaki}},\ and\ \bibinfo {author} {\bibfnamefont {P.}~\bibnamefont
  {Lecheminant}},\ }\href {https://doi.org/10.1103/PhysRevB.90.075116}
  {\bibfield  {journal} {\bibinfo  {journal} {Phys. Rev. B}\ }\textbf {\bibinfo
  {volume} {90}},\ \bibinfo {pages} {075116} (\bibinfo {year}
  {2014})}\BibitemShut {NoStop}%
\bibitem [{\citenamefont {Kang}\ and\ \citenamefont {Moore}(2017)}]{Kang17}%
  \BibitemOpen
  \bibfield  {author} {\bibinfo {author} {\bibfnamefont {B.}~\bibnamefont
  {Kang}}\ and\ \bibinfo {author} {\bibfnamefont {J.~E.}\ \bibnamefont
  {Moore}},\ }\href {https://doi.org/10.1103/PhysRevB.95.245117} {\bibfield
  {journal} {\bibinfo  {journal} {Phys. Rev. B}\ }\textbf {\bibinfo {volume}
  {95}},\ \bibinfo {pages} {245117} (\bibinfo {year} {2017})}\BibitemShut
  {NoStop}%
\bibitem [{\citenamefont {Freedman}\ \emph {et~al.}(2012)\citenamefont
  {Freedman}, \citenamefont {Gukelberger}, \citenamefont {Hastings},
  \citenamefont {Trebst}, \citenamefont {Troyer},\ and\ \citenamefont
  {Wang}}]{Freedman12}%
  \BibitemOpen
  \bibfield  {author} {\bibinfo {author} {\bibfnamefont {M.~H.}\ \bibnamefont
  {Freedman}}, \bibinfo {author} {\bibfnamefont {J.}~\bibnamefont
  {Gukelberger}}, \bibinfo {author} {\bibfnamefont {M.~B.}\ \bibnamefont
  {Hastings}}, \bibinfo {author} {\bibfnamefont {S.}~\bibnamefont {Trebst}},
  \bibinfo {author} {\bibfnamefont {M.}~\bibnamefont {Troyer}},\ and\ \bibinfo
  {author} {\bibfnamefont {Z.}~\bibnamefont {Wang}},\ }\href
  {https://doi.org/10.1103/PhysRevB.85.045414} {\bibfield  {journal} {\bibinfo
  {journal} {Phys. Rev. B}\ }\textbf {\bibinfo {volume} {85}},\ \bibinfo
  {pages} {045414} (\bibinfo {year} {2012})}\BibitemShut {NoStop}%
\bibitem [{\citenamefont {Regnault}\ \emph {et~al.}(2009)\citenamefont
  {Regnault}, \citenamefont {Bernevig},\ and\ \citenamefont
  {Haldane}}]{Regnault09}%
  \BibitemOpen
  \bibfield  {author} {\bibinfo {author} {\bibfnamefont {N.}~\bibnamefont
  {Regnault}}, \bibinfo {author} {\bibfnamefont {B.~A.}\ \bibnamefont
  {Bernevig}},\ and\ \bibinfo {author} {\bibfnamefont {F.~D.~M.}\ \bibnamefont
  {Haldane}},\ }\href {https://doi.org/10.1103/PhysRevLett.103.016801}
  {\bibfield  {journal} {\bibinfo  {journal} {Phys. Rev. Lett.}\ }\textbf
  {\bibinfo {volume} {103}},\ \bibinfo {pages} {016801} (\bibinfo {year}
  {2009})}\BibitemShut {NoStop}%
\bibitem [{\citenamefont {T\ifmmode~\mbox{\H{o}}\else \H{o}\fi{}ke}\ and\
  \citenamefont {Jain}(2009)}]{Toke2009}%
  \BibitemOpen
  \bibfield  {author} {\bibinfo {author} {\bibfnamefont {C.}~\bibnamefont
  {T\ifmmode~\mbox{\H{o}}\else \H{o}\fi{}ke}}\ and\ \bibinfo {author}
  {\bibfnamefont {J.~K.}\ \bibnamefont {Jain}},\ }\href
  {https://doi.org/10.1103/PhysRevB.80.205301} {\bibfield  {journal} {\bibinfo
  {journal} {Phys. Rev. B}\ }\textbf {\bibinfo {volume} {80}},\ \bibinfo
  {pages} {205301} (\bibinfo {year} {2009})}\BibitemShut {NoStop}%
\bibitem [{\citenamefont {Bernevig}\ \emph {et~al.}(2012)\citenamefont
  {Bernevig}, \citenamefont {Bonderson},\ and\ \citenamefont
  {Regnault}}]{Bernevig12}%
  \BibitemOpen
  \bibfield  {author} {\bibinfo {author} {\bibfnamefont {B.~A.}\ \bibnamefont
  {Bernevig}}, \bibinfo {author} {\bibfnamefont {P.}~\bibnamefont
  {Bonderson}},\ and\ \bibinfo {author} {\bibfnamefont {N.}~\bibnamefont
  {Regnault}},\ }\href@noop {} {} (\bibinfo {year} {2012}),\ \Eprint
  {https://arxiv.org/abs/1207.3305} {arXiv:1207.3305 [cond-mat.mes-hall]}
  \BibitemShut {NoStop}%
\bibitem [{\citenamefont {Estienne}\ \emph {et~al.}(2015)\citenamefont
  {Estienne}, \citenamefont {Regnault},\ and\ \citenamefont
  {Bernevig}}]{Estienne15}%
  \BibitemOpen
  \bibfield  {author} {\bibinfo {author} {\bibfnamefont {B.}~\bibnamefont
  {Estienne}}, \bibinfo {author} {\bibfnamefont {N.}~\bibnamefont {Regnault}},\
  and\ \bibinfo {author} {\bibfnamefont {B.~A.}\ \bibnamefont {Bernevig}},\
  }\href {https://doi.org/10.1103/PhysRevLett.114.186801} {\bibfield  {journal}
  {\bibinfo  {journal} {Phys. Rev. Lett.}\ }\textbf {\bibinfo {volume} {114}},\
  \bibinfo {pages} {186801} (\bibinfo {year} {2015})}\BibitemShut {NoStop}%
\bibitem [{\citenamefont {Wu}\ \emph {et~al.}(2014)\citenamefont {Wu},
  \citenamefont {Estienne}, \citenamefont {Regnault},\ and\ \citenamefont
  {Bernevig}}]{Wu2014}%
  \BibitemOpen
  \bibfield  {author} {\bibinfo {author} {\bibfnamefont {Y.-L.}\ \bibnamefont
  {Wu}}, \bibinfo {author} {\bibfnamefont {B.}~\bibnamefont {Estienne}},
  \bibinfo {author} {\bibfnamefont {N.}~\bibnamefont {Regnault}},\ and\
  \bibinfo {author} {\bibfnamefont {B.~A.}\ \bibnamefont {Bernevig}},\ }\href
  {https://doi.org/10.1103/PhysRevLett.113.116801} {\bibfield  {journal}
  {\bibinfo  {journal} {Phys. Rev. Lett.}\ }\textbf {\bibinfo {volume} {113}},\
  \bibinfo {pages} {116801} (\bibinfo {year} {2014})}\BibitemShut {NoStop}%
\bibitem [{\citenamefont {Milovanovi\ifmmode~\acute{c}\else \'{c}\fi{}}\ \emph
  {et~al.}(2009)\citenamefont {Milovanovi\ifmmode~\acute{c}\else \'{c}\fi{}},
  \citenamefont {Jolicoeur},\ and\ \citenamefont
  {Vidanovi\ifmmode~\acute{c}\else \'{c}\fi{}}}]{Milovanovic2009}%
  \BibitemOpen
  \bibfield  {author} {\bibinfo {author} {\bibfnamefont {M.~V.}\ \bibnamefont
  {Milovanovi\ifmmode~\acute{c}\else \'{c}\fi{}}}, \bibinfo {author}
  {\bibfnamefont {T.}~\bibnamefont {Jolicoeur}},\ and\ \bibinfo {author}
  {\bibfnamefont {I.}~\bibnamefont {Vidanovi\ifmmode~\acute{c}\else
  \'{c}\fi{}}},\ }\href {https://doi.org/10.1103/PhysRevB.80.155324} {\bibfield
   {journal} {\bibinfo  {journal} {Phys. Rev. B}\ }\textbf {\bibinfo {volume}
  {80}},\ \bibinfo {pages} {155324} (\bibinfo {year} {2009})}\BibitemShut
  {NoStop}%
\bibitem [{\citenamefont {Yang}\ \emph {et~al.}(2019)\citenamefont {Yang},
  \citenamefont {Wu},\ and\ \citenamefont {Papi\ifmmode~\acute{c}\else
  \'{c}\fi{}}}]{BoYang19}%
  \BibitemOpen
  \bibfield  {author} {\bibinfo {author} {\bibfnamefont {B.}~\bibnamefont
  {Yang}}, \bibinfo {author} {\bibfnamefont {Y.-H.}\ \bibnamefont {Wu}},\ and\
  \bibinfo {author} {\bibfnamefont {Z.}~\bibnamefont
  {Papi\ifmmode~\acute{c}\else \'{c}\fi{}}},\ }\href
  {https://doi.org/10.1103/PhysRevB.100.245303} {\bibfield  {journal} {\bibinfo
   {journal} {Phys. Rev. B}\ }\textbf {\bibinfo {volume} {100}},\ \bibinfo
  {pages} {245303} (\bibinfo {year} {2019})}\BibitemShut {NoStop}%
\bibitem [{\citenamefont {Yang}(2021)}]{BoYang21}%
  \BibitemOpen
  \bibfield  {author} {\bibinfo {author} {\bibfnamefont {B.}~\bibnamefont
  {Yang}},\ }\href {https://doi.org/10.1103/PhysRevB.103.115102} {\bibfield
  {journal} {\bibinfo  {journal} {Phys. Rev. B}\ }\textbf {\bibinfo {volume}
  {103}},\ \bibinfo {pages} {115102} (\bibinfo {year} {2021})}\BibitemShut
  {NoStop}%
\bibitem [{\citenamefont {Mandal}(2018)}]{Mandal18_general}%
  \BibitemOpen
  \bibfield  {author} {\bibinfo {author} {\bibfnamefont {S.~S.}\ \bibnamefont
  {Mandal}},\ }\href {https://doi.org/10.1088/1361-648x/aadd37} {\bibfield
  {journal} {\bibinfo  {journal} {Journal of Physics: Condensed Matter}\
  }\textbf {\bibinfo {volume} {30}},\ \bibinfo {pages} {405605} (\bibinfo
  {year} {2018})}\BibitemShut {NoStop}%
\bibitem [{\citenamefont {Haldane}(1983)}]{HaldaneSphere}%
  \BibitemOpen
  \bibfield  {author} {\bibinfo {author} {\bibfnamefont {F.~D.~M.}\
  \bibnamefont {Haldane}},\ }\href {https://doi.org/10.1103/PhysRevLett.51.605}
  {\bibfield  {journal} {\bibinfo  {journal} {Phys. Rev. Lett.}\ }\textbf
  {\bibinfo {volume} {51}},\ \bibinfo {pages} {605} (\bibinfo {year}
  {1983})}\BibitemShut {NoStop}%
\bibitem [{\citenamefont {Das}\ \emph {et~al.}(2021)\citenamefont {Das},
  \citenamefont {Das},\ and\ \citenamefont {Mandal}}]{Das21}%
  \BibitemOpen
  \bibfield  {author} {\bibinfo {author} {\bibfnamefont {S.}~\bibnamefont
  {Das}}, \bibinfo {author} {\bibfnamefont {S.}~\bibnamefont {Das}},\ and\
  \bibinfo {author} {\bibfnamefont {S.~S.}\ \bibnamefont {Mandal}},\ }\href
  {https://doi.org/10.1103/PhysRevB.103.075304} {\bibfield  {journal} {\bibinfo
   {journal} {Phys. Rev. B}\ }\textbf {\bibinfo {volume} {103}},\ \bibinfo
  {pages} {075304} (\bibinfo {year} {2021})}\BibitemShut {NoStop}%
\bibitem [{\citenamefont {Li}\ and\ \citenamefont
  {Haldane}(2008)}]{LiHaldane2008}%
  \BibitemOpen
  \bibfield  {author} {\bibinfo {author} {\bibfnamefont {H.}~\bibnamefont
  {Li}}\ and\ \bibinfo {author} {\bibfnamefont {F.~D.~M.}\ \bibnamefont
  {Haldane}},\ }\href {https://doi.org/10.1103/PhysRevLett.101.010504}
  {\bibfield  {journal} {\bibinfo  {journal} {Phys. Rev. Lett.}\ }\textbf
  {\bibinfo {volume} {101}},\ \bibinfo {pages} {010504} (\bibinfo {year}
  {2008})}\BibitemShut {NoStop}%
\bibitem [{\citenamefont {Rodr\'{\i}guez}\ \emph {et~al.}(2012)\citenamefont
  {Rodr\'{\i}guez}, \citenamefont {Simon},\ and\ \citenamefont
  {Slingerland}}]{Simon12}%
  \BibitemOpen
  \bibfield  {author} {\bibinfo {author} {\bibfnamefont {I.~D.}\ \bibnamefont
  {Rodr\'{\i}guez}}, \bibinfo {author} {\bibfnamefont {S.~H.}\ \bibnamefont
  {Simon}},\ and\ \bibinfo {author} {\bibfnamefont {J.~K.}\ \bibnamefont
  {Slingerland}},\ }\href {https://doi.org/10.1103/PhysRevLett.108.256806}
  {\bibfield  {journal} {\bibinfo  {journal} {Phys. Rev. Lett.}\ }\textbf
  {\bibinfo {volume} {108}},\ \bibinfo {pages} {256806} (\bibinfo {year}
  {2012})}\BibitemShut {NoStop}%
\bibitem [{\citenamefont {Sterdyniak}\ \emph {et~al.}(2012)\citenamefont
  {Sterdyniak}, \citenamefont {Chandran}, \citenamefont {Regnault},
  \citenamefont {Bernevig},\ and\ \citenamefont {Bonderson}}]{Sterdyniak12}%
  \BibitemOpen
  \bibfield  {author} {\bibinfo {author} {\bibfnamefont {A.}~\bibnamefont
  {Sterdyniak}}, \bibinfo {author} {\bibfnamefont {A.}~\bibnamefont
  {Chandran}}, \bibinfo {author} {\bibfnamefont {N.}~\bibnamefont {Regnault}},
  \bibinfo {author} {\bibfnamefont {B.~A.}\ \bibnamefont {Bernevig}},\ and\
  \bibinfo {author} {\bibfnamefont {P.}~\bibnamefont {Bonderson}},\ }\href
  {https://doi.org/10.1103/PhysRevB.85.125308} {\bibfield  {journal} {\bibinfo
  {journal} {Phys. Rev. B}\ }\textbf {\bibinfo {volume} {85}},\ \bibinfo
  {pages} {125308} (\bibinfo {year} {2012})}\BibitemShut {NoStop}%
\bibitem [{\citenamefont {Banerjee}\ \emph {et~al.}(2018)\citenamefont
  {Banerjee}, \citenamefont {Heiblum}, \citenamefont {Umansky}, \citenamefont
  {Feldman}, \citenamefont {Oreg},\ and\ \citenamefont {Stern}}]{Banerjee2018}%
  \BibitemOpen
  \bibfield  {author} {\bibinfo {author} {\bibfnamefont {M.}~\bibnamefont
  {Banerjee}}, \bibinfo {author} {\bibfnamefont {M.}~\bibnamefont {Heiblum}},
  \bibinfo {author} {\bibfnamefont {V.}~\bibnamefont {Umansky}}, \bibinfo
  {author} {\bibfnamefont {D.~E.}\ \bibnamefont {Feldman}}, \bibinfo {author}
  {\bibfnamefont {Y.}~\bibnamefont {Oreg}},\ and\ \bibinfo {author}
  {\bibfnamefont {A.}~\bibnamefont {Stern}},\ }\href
  {https://doi.org/10.1038/s41586-018-0184-1} {\bibfield  {journal} {\bibinfo
  {journal} {Nature}\ }\textbf {\bibinfo {volume} {559}},\ \bibinfo {pages}
  {205} (\bibinfo {year} {2018})}\BibitemShut {NoStop}%
\bibitem [{\citenamefont {Feldman}\ and\ \citenamefont
  {Halperin}(2021)}]{Halperin21}%
  \BibitemOpen
  \bibfield  {author} {\bibinfo {author} {\bibfnamefont {D.~E.}\ \bibnamefont
  {Feldman}}\ and\ \bibinfo {author} {\bibfnamefont {B.~I.}\ \bibnamefont
  {Halperin}},\ }\href {https://doi.org/10.1088/1361-6633/ac03aa} {\bibfield
  {journal} {\bibinfo  {journal} {Rep. Prog. Phys.}\ }\textbf {\bibinfo
  {volume} {84}},\ \bibinfo {pages} {076501} (\bibinfo {year}
  {2021})}\BibitemShut {NoStop}%
\end{thebibliography}%

\end{document}